\documentclass[preprint]{aastex}
\usepackage{epsf}
\newcommand{\simgt}{\lower.5ex\hbox{$\; \buildrel > \over \sim \;$}}
\newcommand{\simlt}{\lower.5ex\hbox{$\; \buildrel < \over \sim \;$}}
\newcommand{\himpc}{{\hbox {$h^{-1}$}{\rm Mpc}} }
\newcommand{\deltah}{{\delta_{\rm halo}}}
\newcommand{\deltam}{{\delta_{\rm mass}}}
\newcommand{\zf}{{z_{\rm f}}}
\newcommand{\halo}{{\Delta_{\rm h}}}
\newcommand{\sigmamm}{{\sigma_{\rm mm}}}

\newcommand{\sigmahh}{{\sigma_{\rm hh}}}
\newcommand{\Rs}{{R_{\rm\scriptscriptstyle S}}}
\newcommand{\deltaRG}{{\delta_{\rm\scriptscriptstyle RG}}}
\newcommand{\gmax}{{g_{\rm\scriptscriptstyle MAX}}}
\slugcomment{The Astrophysical Journal (2001), in press. UTAP-380/2000}
\shorttitle{Genus Statistic for Galaxy Clusters}
\shortauthors{Hikage, Taruya \& Suto}
\begin{document}
%
%%%%%%%%%%%%%%%%%%%%%%%%%%%%%%%%%%%%%%%%%%%%%%%%%%%%%%%%%%%%%%%%%%%%%%
%%%%%%%%%%%%%%%%%%%%%%%%%%%%%%%%%%%%%%%%%%%%%%%%%%%%%%%%%%%%%%%%%%%%%%
\title{Genus statistics for galaxy clusters \\
and nonlinear biasing of dark matter halos}
%%%%%%%%%%%%%%%%%%%%%%%%%%%%%%%%%%%%%%%%%%%%%%%%%%%%%%%%%%%%%%%%%%%%%%
%%%%%%%%%%%%%%%%%%%%%%%%%%%%%%%%%%%%%%%%%%%%%%%%%%%%%%%%%%%%%%%%%%%%%%
%
%%%%%%%%%%%%%%%%%%%%%%%%%%%%%%%%%%%%%%%%%%%%%%%%%%%%%%%%%%%%%%%%%%%%%%
\author{Chiaki Hikage, Atsushi Taruya and Yasushi Suto\footnote{Also
at  Research Center for
    the Early Universe (RESCEU), School of Science, University of
    Tokyo, Tokyo 113-0033, Japan.}}
\affil{Department of Physics, School of Science, University of
    Tokyo, Tokyo 113-0033, Japan.}
\email{hikage@utap.phys.s.u-tokyo.ac.jp,
    ataruya@utap.phys.s.u-tokyo.ac.jp, suto@phys.s.u-tokyo.ac.jp}
%%%%%%%%%%%%%%%%%%%%%%%%%%%%%%%%%%%%%%%%%%%%%%%%%%%%%%%%%%%%%%%%%%%%%%
%
\received{2000 November 24}
\accepted{2001 ??}
\begin{abstract}
  We derive an analytic formula of genus for dark matter halos
  assuming that the primordial mass density field obeys the
  random-Gaussian statistics.  In particular, we for the first time
  take account of the nonlinear nature of the halo
  biasing, in addition to the nonlinear gravitational evolution of the
  underlying mass distribution. We examine in detail the extent to
  which the predicted genus for halos depends on the redshift,
  smoothing scale of the density field, and the mass of those halos in
  representative cold dark matter models. In addition to the full
  model predictions, we derive an explicit perturbation formula for
  the halo genus which can be applied to a wider class of biasing
  model for dark matter halos.  With the empirical relation between
  the halo mass and the luminosity/temperature of galaxy clusters, our
  prediction can be compared with the upcoming cluster redshift
  catalogues in optical and X-ray bands.  
  Furthermore, we discuss the detectability of
  the biasing and cosmological model dependence from the future data.
\end{abstract}
%
%%%%%%%%%%%%%%%%%%%%%%%%%%%%%%%%%%%%%%%%%%%%%%%%%%%%%%%%%%%
%%%%%%%%%%%%%%%%%%%%%%%%%%%%%%%%%%%%%%%%%%%%%%%%%%%%%%%%%%%
\newpage

\section{INTRODUCTION}
\label{sec:intro}

The genus statistics is a quantitative measure of the topological nature
of the density field. In contrast to the more conventional two-point
statistics like the two-point correlation functions and the power
spectrum, genus is sensitive to the phase information of the density
field, and therefore plays a complementary role in characterizing the
present cosmic structure.

Mathematically, genus $G(\nu)$ is defined as $-1/2$ times the Euler
characteristic of the isodensity contour of the density field at the
level of $\nu$ times the rms fluctuations $\sigma$.  In practice this is
equal to (number of holes) $-$ (number of isolated regions) of the
isodensity surface.  In cosmology it is convenient to use the genus per
unit volume, or {\it the genus density} $g(\nu)$, rather than the genus
itself.  The genus density of the random-Gaussian density field is
analytically given by
%%%%%%%%%%%%%%%%%%%%%%%%%%%%%%%%%%%%%%%%%%%%%%%%%%%%%%%%%%%%%%
\begin{equation}
\label{eq:genus_rd}
g(\nu) = \frac{1}{(2\pi)^2}\left(\frac{\sigma_1^2}{3\sigma^2}
\right)^{3/2}(1-\nu^2)\exp\left(-\frac{\nu^2}{2}\right),
\end{equation}
%%%%%%%%%%%%%%%%%%%%%%%%%%%%%%%%%%%%%%%%%%%%%%%%%%%%%%%%%%%%%%
where $\sigma_1 \equiv \langle[\nabla\delta(x)]^2\rangle$ and $\sigma
\equiv \langle[\delta(x)]^2\rangle$ with $\langle \rangle$ denoting the
average over the probability distribution function (PDF) of the density
field (Doroshkevich 1970; Adler 1981; 
Bardeen et al. 1986; Gott, Melott, \& Dickinson 1986; 
Hamilton, Gott, \& Weinberg 1986). 

In fact the standard model of structure formation assumes that the
primordial density field responsible for the current cosmic structure is
random-Gaussian. Thus equation (\ref{eq:genus_rd}) is the most important
analytical result for genus, and has been compared with the
observational estimates from galaxy and cluster samples in order to
probe the random-Gaussianity in the primordial density field (e.g.,
Vogeley et al. 1994; Rhoads et al. 1994; Canavezes et al. 2000).
Nevertheless the nonlinear gravitational evolution inevitably changes
the random-Gaussianity.  Matsubara (1994) explicitly finds the
next-order terms for the genus using the perturbation analysis of the
evolution of the density field.  His formula is in good agreement with
the results from N-body simulations (Matsubara \& Suto 1996; Colley et
al. 2000) in a weakly nonlinear regime.  The genus in a strongly
nonlinear regime has been considered using a variety of modelings
including the direct numerical simulations (Matsubara \& Suto 1996), the
empirical log-normal PDF for cosmological nonlinear density field
(Matsubara \& Yokoyama 1996), and the Zel'dovich approximation (Seto et
al. 1997).

Those previous studies, however, did not take account of the effect of
biasing between luminous objects and dark matter.  On the other hand,
there is growing evidence that the biasing is described by a nonlinear
and stochastic function of the underlying mass density (Dekel \& Lahav
1999).  If this is the case, the genus for the isodensity surface of
luminous objects should be significantly different from that of mass.
Of course the biasing is supposed to be very dependent on the specific
objects, and the general prediction is virtually impossible. Therefore
in this paper we focus on a nonlinear stochastic biasing model of dark
matter halos (Taruya \& Suto 2000). This model is based on the extended
Press-Schechter theory (e.g., Bower 1991; Bond et al. 1991) combined
with the Mo \& White (1996) bias formula, and is in reasonable agreement
with numerical simulations (Kravtsov \& Klypin 1999; Somerville et
al. 2001; Yoshikawa et al. 2001).  In this paper, we present an analytic
expression for genus of dark matter halos by taking account of the
nonlinearity in the halo biasing.  Since recent numerical simulations
(Taruya et al. 2001) suggest the stochasticity in halo biasing is fairly
small on scales $\simgt 10 h^{-1}$Mpc (see their Fig.8), we neglect the
effect of stochasticity in what follows.  The one-to-one correspondence
between those halos and galaxy clusters is a fairly standard assumption,
and our predictions for halo genus as a function of halo mass can be
translated into those of clusters with the corresponding
luminosity/temperature. Therefore our predictions are useful in future
data analysis of the cluster surveys in different bands.

The plan of this paper is as follows.  In \S 2, we briefly review
the nonlinear stochastic biasing model for dark matter halos that we
adopt, and derive an analytic expression of genus for halos.  In \S 3,
we present specific predictions of genus statistics paying particular
attention to the dependence on the smoothing length, the range of halo
mass, cosmological parameters and the redshift.  Finally \S 4 is devoted
to conclusions and discusses implications of our results.

%%%%%%%%%%%%%%%%%%%%%%%%%%%%%%%%%%%%%%%%%%%%%%%%%%%%%%%%%%%%%%%

\section{PREDICTING THE GENUS FOR HALOS \label{sec:pretict}}

In this section we derive an analytic expression for genus of halos
adopting the nonlinear stochastic biasing model (Taruya \& Suto 2000).
First we briefly review the halo biasing model (\S 2.1).  Assuming
that PDF of nonlinear mass fluctuations is given by the log-normal
distribution, we compute the PDF of halo number density fluctuations,
$P_{\rm halo}(\deltah)$ (\S 2.2).  Finally in \S 2.3 we present our
analytic expression for genus of halos which properly takes account of
both the biasing and the nonlinear gravitational evolution for the
first time.

\subsection{Halo Biasing Model \label{subsec:biasing}}

The central idea of the halo biasing model of Taruya \& Suto (2000) is
to regard the formation epoch and mass of halos as the major two {\it
hidden} (unobservable) variables which lead to the nonlinear stochastic
behavior of the halo number density field, $\deltah$, as a function of
the underlying mass density fluctuation, $\deltam$.  Applying the
extended Press-Schechter theory, Mo \& White (1996) derived an
expression of $\deltah$ at $z$ smoothed over the scale of $\Rs$ as a
function of $\deltam$, the halo mass $M$, and its formation epoch
$\zf$:
%%%%%%%%%%%%%%%%%%%%%%%%%%%%%%%%%%%%%%%%%%%%%%%%%%%%%%%%%%%%%
\begin{equation}
\deltah = \halo(\Rs,z|\deltam,M,\zf) .
\end{equation}
%%%%%%%%%%%%%%%%%%%%%%%%%%%%%%%%%%%%%%%%%%%%%%%%%%%%%%%%%%%%%
Since the above procedure assumes the spherical collapse model, we adopt
the top-hat smoothing in computing $\deltam$ and $\deltah$ throughout
the paper.  Convolving the above expression with the Press--Schechter
mass function $n(M,z;\delta_{\rm c,0})$ and the halo formation epoch
distribution function, $\partial p/\partial \zf(\zf | M,z)$ (Lacey
\& Cole 1993; Kitayama \& Suto 1996), Taruya \& Suto (2000) derive the
following conditional PDF of $\deltah$ for a given $\deltam$,
$P(\deltah|\deltam)$:
%%%%%%%%%%%%%%%%%%%%%%%%%%%%%%%%%%%%%%%%%%%%%%%%%%%%%%%%%%%%%%%%%%
\begin{equation}
    P(\deltah|\deltam) ~d\deltah~=~{\cal N}^{-1}~
    \int\int_{{\cal C}(M,\zf)}dM~ d\zf~~
\frac{\partial p}{\partial \zf}(\zf | M,z)~n(M,z;\delta_{\rm c,0}) ,
\label{weight}
\end{equation}
%%%%%%%%%%%%%%%%%%%%%%%%%%%%%%%%%%%%%%%%%%%%%%%%%%%%%%%%%%%%%%%%%%
where $\delta_{\rm c,0}$ is the critical threshold for spherical
collapse $3(12\pi)^{3/2}/20 \simeq 1.69$.  The region of the
integration, ${\cal C}(M,\zf)$, is given as follows:
%%%%%%%%%%%%%%%%%%%%%%%%%%%%%%%%%%%%%%%%%%%%%%%%%%%%%%%%%%%%%%%%%%%%%%%%
\begin{eqnarray}
    \label{eq:constraint}
{\cal C}(M,\zf) = \{~ (M,\zf)~|~
     \deltah\leq~\halo(\Rs,z|\deltam,M,\zf)~\leq\deltah+d\deltah , \cr
 M_{\rm min}\leq M \leq M_{\rm max}, \quad z\leq \zf \leq \infty ~\} ,
\end{eqnarray}
%%%%%%%%%%%%%%%%%%%%%%%%%%%%%%%%%%%%%%%%%%%%%%%%%%%%%%%%%%%%%%%%%%
where $M_{\rm min}$ and $M_{\rm max}$ denote maximum and minimum of
halo mass.  The normalization factor ${\cal N}$ is defined as
%%%%%%%%%%%%%%%%%%%%%%%%%%%%%%%%%%%%%%%%%%%%%%%%%%%%%%%%%%%%%%%%%%%%%
\begin{equation}
    \label{eq:normalization}
{\cal N}=
    \int_{M_{\rm min}}^{M_{\rm max}}dM \int_{z}^{\infty} d\zf~~
    \frac{\partial p}{\partial \zf}(\zf|M,z)~n(M,z;\delta_{\rm c,0}) .
\end{equation}
%%%%%%%%%%%%%%%%%%%%%%%%%%%%%%%%%%%%%%%%%%%%%%%%%%%%%%%%%%%%%%%%%%%%%
The joint PDF $P(\deltam,\deltah)$ is simply given by multiplying
the PDF of $\deltam$, $P(\deltam)$, which we assume is log-normal:
%%%%%%%%%%%%%%%%%%%%%%%%%%%%%%%%%%%%%%%%%%%%%%%%%%%%%%%%%%%%%%%%%%%%%
\begin{equation}
\label{eq:jointpdf}
P(\deltam,\deltah) = P(\deltah|\deltam) \, P(\deltam) .
\end{equation}
%%%%%%%%%%%%%%%%%%%%%%%%%%%%%%%%%%%%%%%%%%%%%%%%%%%%%%%%%%%%%%%%%%%%%

It is well known that the PDF of the nonlinear mass density field is
approximated by the log-normal distribution (e.g. Coles \& Jones 1991;
Kofman et al. 1994). Using this PDF in the present context, however,
implies that we implicitly assume the one-to-one correspondence of the
primordial density fluctuation and the evolved nonlinear one
($\deltam$).  Recently Kayo et al. (2001) show that the scatter around
the mean relation between the initial and final density fluctuations
is significant although the nonlinear PDF is in good agreement with
the log-normal. Thus it should be noted that the present paper
neglects the effect of the scatter or the stochasticity of the mass
density field. This effect will be examined elsewhere when we compare
the predictions of this paper with results of N-body simulations
(Hikage et al. in preparation). In the same spirit, we below do not
take into account the stochasticity of $\deltah$ with respect to
$\deltam$, and adopt the one-to-one mapping between $\deltah$ and
$\deltam$ represented by the mean biasing:
%%%%%%%%%%%%%%%%%%%%%%%%%%%%%%%%%%%%%%%%%%%%%%%%%%%%%%%%%%%%%%%%%%%%%
\begin{equation}
    \label{eq:mean_biasing}
    \overline{\deltah} = \int P(\deltah|\deltam)~\deltah~d\deltah~
    \equiv  f_{\rm bias}(\deltam) .
\end{equation}
%%%%%%%%%%%%%%%%%%%%%%%%%%%%%%%%%%%%%%%%%%%%%%%%%%%%%%%%%%%%%%%%%%%%%
In what follows, we use $\deltah$ to indicate
$\overline{\deltah}(\deltam)= f_{\rm bias}(\deltam)$ in reality.

In \S 3.2, we show that our results can be well reproduced from the mean
biasing function up to the second-order term of $\deltam$.  In fact,
this is important in the sense that our genus expression in the next
section is applicable to the more general biasing scheme as long as its
second-order parameterization form is specified.  In the present biasing
model, we first compute the mean biasing numerically according to
equation (\ref{eq:mean_biasing}), and then fit the data in the range of
$-0.1<\deltam<0.1$ to the following quadratic model:
%%%%%%%%%%%%%%%%%%%%%%%%%%%%%%%%%%%%%%%%%%%%%%%%%%%%%%%%%%%%%%%%%%%
\begin{equation}
\label{eq:biasing_fit}
   f_{\rm bias}^{\rm fit(2)}(\deltam) = b_1\deltam 
   + \frac{b_2}{2}(\deltam^2 - \sigmamm^2) 
\end{equation}
%%%%%%%%%%%%%%%%%%%%%%%%%%%%%%%%%%%%%%%%%%%%%%%%%%%%%%%%%%%%%%%%%%%
which satisfy the relation $\langle f_{\rm bias}^{\rm
fit(2)}(\deltam)\rangle = 0$ ($\sigmamm$ is defined in
eq.[\ref{eq:peacock-dodds}] below).  The fitted values for the linear
and the second-order biasing coefficients, $b_{\rm 1}$ and $b_{\rm 2}$,
are listed in Table \ref{tab:biasing_fit} for several models.

\begin{figure}
\begin{center}
   \leavevmode \epsfxsize=16cm \epsfbox{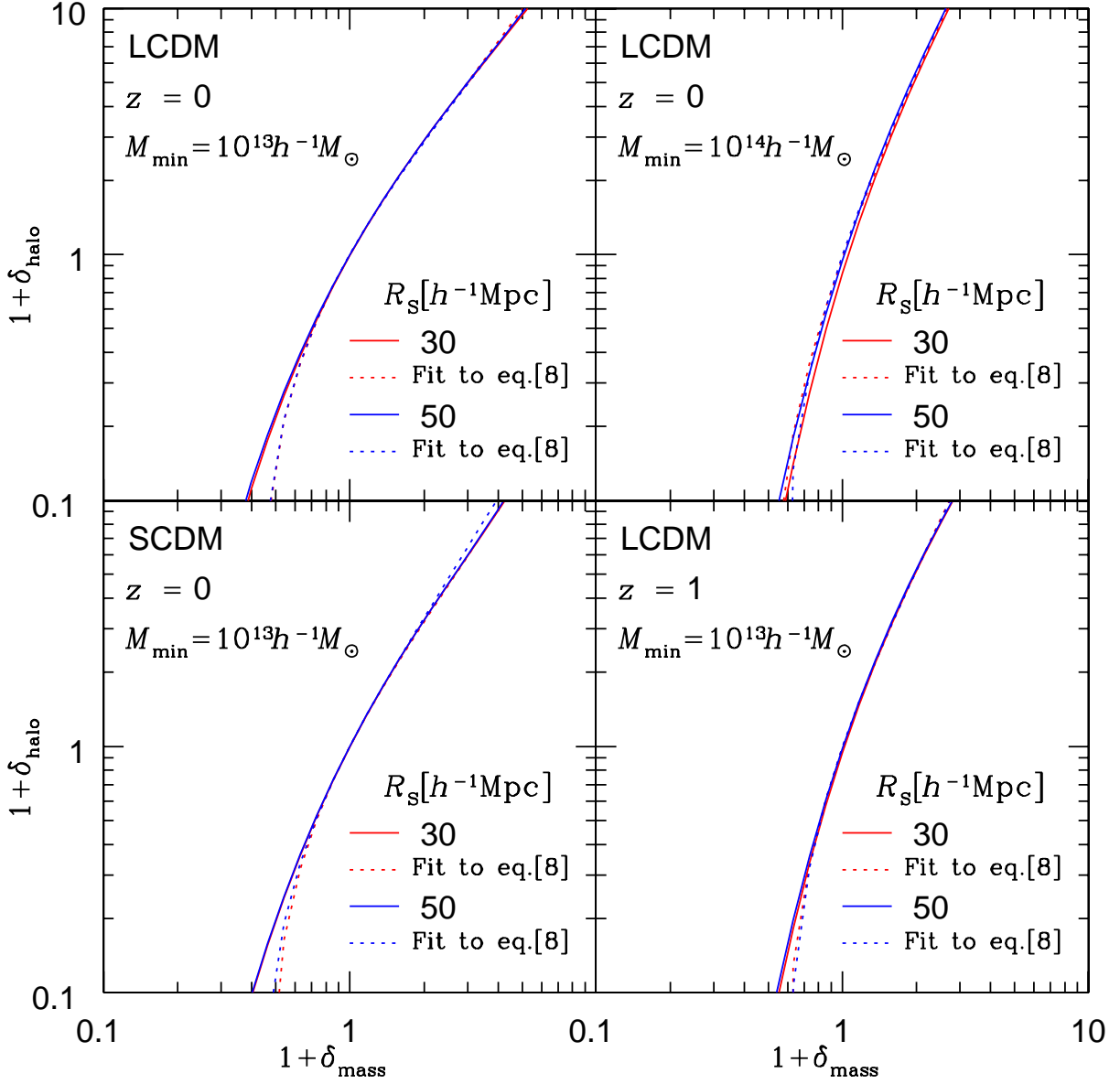}
\end{center}
\figcaption{The mean biasing function $f_{\rm bias}(\deltam)$({\it
solid}) and its fitting up to the second-order of $\deltam$, $f_{\rm
bias}^{\rm fit(2)}(\deltam)$({\it dotted}).  The red and blue lines
indicate the results with $\Rs=30h^{-1}$Mpc and $50h^{-1}$Mpc.
{\it Upper-left:} LCDM model, $z=0$, and $M_{\rm min} =
10^{13}h^{-1}M_{\odot}$ ; {\it Upper-right:} LCDM model, $z=0$, and
$M_{\rm min} = 10^{14}h^{-1}M_{\odot}$ ; {\it Lower-left:} SCDM model,
$z=0$, and $M_{\rm min} = 10^{13}h^{-1}M_{\odot}$ ; {\it Lower-right:}
LCDM model, $z=1$, and $M_{\rm min} = 10^{13}h^{-1}M_{\odot}$.  We set
$M_{\rm max} = 10^{16}h^{-1}M_{\odot}$.
\label{fig:biasing} }
\end{figure}

Figure \ref{fig:biasing} plots $f_{\rm bias}(\deltam)$ and $f_{\rm
  bias}^{\rm fit(2)}(\deltam)$ in solid and dotted lines, for
different $M_{\rm min}$, $\Rs$, and $z$.
Since we are primarily interested in the galaxy cluster scales, we
consider the following values of the parameters; the minimum mass of halo
$M_{\rm min} = 10^{13}h^{-1}M_{\odot}$ and $10^{14}h^{-1}M_{\odot}$,
the smoothing length $\Rs= 30h^{-1}{\rm Mpc}$ and $50h^{-1}{\rm
  Mpc}$, $z=0$ and $1$ ($h$ denotes the Hubble constant in units of
100 km$\cdot$s$^{-1} \cdot$Mpc$^{-1}$).  While we set the maximum halo
mass $M_{\rm max}=10^{16}h^{-1}M_{\odot}$, the larger value of $M_{\rm
  max}$ does not change the results below since the number density of
such massive halos is quite small. Our predictions below are computed
in two representative Cold Dark Matter (CDM) models; Lambda CDM (LCDM)
with ($\Omega_0,\lambda_0,\sigma_8,h$) = (0.3,0.7,1.0,0.7), and
Standard CDM (SCDM) with
($\Omega_0,\lambda_0,\sigma_8,h$)=(1.0,0.0,0.6,0.5).  The density
parameter $\Omega_0$, dimensionless cosmological constant,
$\lambda_0$, and the top-hat mass fluctuation at $8h^{-1}$Mpc,
$\sigma_8$ are normalized according to the cluster abundance (Kitayama
\& Suto 1997).

Figure \ref{fig:biasing} clearly indicates the degree of nonlinearity
in our biasing model. For the range of parameter values of our
interest, the quadratic fit $f_{\rm bias}^{\rm fit(2)}$ provides a
reasonable approximation to the mean biasing. As $M_{\rm min}$ and $z$
increase, nonlinearity in the biasing becomes stronger (see $b_{\rm
  2}$ in Table \ref{tab:biasing_fit} ), and one expects the more
significant departure of the resulting genus density from the
random-Gaussian prediction (eq.[\ref{eq:genus_rd}]) as we will show
below.

\subsection{Probability Distribution Function of Nonlinear Mass and Halo
  Density Fields \label{subsec:pdf}}

PDF of the objects of interest is the most important and basic
statistics in computing their genus. The PDF of halos in our biasing
model is derived in this subsection.

Standard models of structure formation inspired by the inflation
picture assume that the PDF of the primordial density field obeys the
random-Gaussian statistics. While the nonlinear gravitational
evolution of the density field distorts the primordial
random-Gaussianity, the log-normal PDF is known to be an empirical
good approximation to the nonlinear mass density field $\deltam$ (see
discussion in the previous subsection):
%%%%%%%%%%%%%%%%%%%%%%%%%%%%%%%%%%%%%%%%%%%%%%%%%%%%%%%%%
\begin{equation}
\label{eq:pdf_mass}
 P_{\rm mass}(\deltam) d\deltam = \frac{1}{(1
 +\deltam)\sqrt{2\pi
 \ln(1+\sigmamm^{2})}}\exp\left[-\frac{\{\ln[(1+\deltam)
 \sqrt{1+\sigmamm ^{2}}]\}^{2}}{2\ln(1+\sigmamm^{2})}\right]
  d\deltam .
\end{equation}
%%%%%%%%%%%%%%%%%%%%%%%%%%%%%%%%%%%%%%%%%%%%%%%%%%%%%%%%%
The variance $\sigmamm(\Rs,z)$ of the mass fluctuations at $z$
smoothed over the scale $\Rs$ is computed from the nonlinear power
spectrum $P(k,z)$ convolved with the specific window function
$W^2(k\Rs)$:
%%%%%%%%%%%%%%%%%%%%%%%%%%%%%%%%%%%%%%%%%%%%%%%%%%%%%%%%%
\begin{equation}
 \label{eq:peacock-dodds}
  \sigmamm^{2}(\Rs,z)=\int\frac{dk}{2\pi^{2}}
   k^{2}P_{\rm nl}(k,z)W^{2}(k\Rs)~ .
\end{equation}
%%%%%%%%%%%%%%%%%%%%%%%%%%%%%%%%%%%%%%%%%%%%%%%%%%%%%%%   
In what follows, we adopt the fitting formula of the nonlinear CDM power
spectrum $P_{\rm nl}(k,z)$ (Peacock \& Dodds 1996).  Throughout the
paper, we adopt the top-hat window function:
%%%%%%%%%%%%%%%%%%%%%%%%%%%%%%%%%%%%%%%%%%%%%%%%%
\begin{eqnarray}
\label{top-hat}
 W(x)= \frac{3}{x^3}(\sin{x}-x\cos{x}) .
\end{eqnarray}
%%%%%%%%%%%%%%%%%%%%%%%%%%%%%%%%%%%%%%%%%%%%%%%%%
The resulting PDF in this prescription proves to be in excellent
agreement with N-body simulation results (Kayo et al. 2001).

One interpretation of equation (\ref{eq:pdf_mass}) is that the nonlinear
mass density field $\deltam$ has the following one-to-one mapping from a
random-Gaussian field $\deltaRG$ with unit variance:
%%%%%%%%%%%%%%%%%%%%%%%%%%%%%%%%%%%%%%%%%%%%%%%%%%%%%%%%%%%%%%%%%%%
\begin{equation}
\label{fig:lognormal-mapping}
\deltam =f_{\rm ln}[\deltaRG] 
= \frac{1}{\sqrt{1+\sigmamm^{2}}}
~\exp~ [\sqrt{\ln(1+\sigmamm^{2})}\deltaRG] - 1 .
\end{equation}
%%%%%%%%%%%%%%%%%%%%%%%%%%%%%%%%%%%%%%%%%%%%%%%%%%%%%%%%%%%%%%%%%%%
Physically speaking, the above $\deltaRG$ corresponds to the
primordial density field apart from the arbitrary normalization factor.
Of course this mapping cannot be strict since $\deltam$ should not be
determined locally. Nevertheless this interpretation provides an
interesting and useful method to compute genus of nonlinear mass
density field as discussed by Matsubara \& Yokoyama (1996).

We extend this idea in our biasing model.  Since our halo density
field is given by the mean biasing (\ref{eq:mean_biasing}) and we
adopt the log-normal PDF (\ref{eq:pdf_mass}) for $\deltam$, we can
easily obtain the PDF for $\deltah=f_{\rm bias}(\deltam)=f_{\rm
  bias}(f_{\rm ln}(\deltaRG))$ as follows:
%%%%%%%%%%%%%%%%%%%%%%%%%%%%%%%%%%%%%%%%%%%%%%%%%%%%
\begin{eqnarray}
\label{eq:pdf_halo}
P_{\rm halo}(\deltah) d\deltah & = & P_{\rm mass}[f^{-1}_{\rm
  bias}(\deltah)] \left. {d \deltam \over d\, f_{\rm bias}(\deltam)}
\right|_{\deltam=f^{-1}_{\rm bias}(\deltah)} d\deltah \\
& =  & \frac{1}{f^{\prime}_{\rm bias}(f_{\rm bias}^{-1}(\deltah))
  (1+f_{\rm bias}^{-1}(\deltah))
  \sqrt{2\pi \ln(1+\sigmamm^{2})}} \nonumber \\
  & \times & \exp\left[-\frac{\{\ln[(1+f_{\rm bias}^{-1}(\deltah))
  \sqrt{1+\sigmamm^{2}}]\}^{2}}{2\ln(1+\sigmamm^{2})}\right]
  d\deltah .
\end{eqnarray}
%%%%%%%%%%%%%%%%%%%%%%%%%%%%%%%%%%%%%%%%%%%%%%%%%%%%%%%%%%%%%%
In the above, $f^{\prime}_{\rm bias}(\deltam)$ denotes the derivative
respect to $\deltam$, and $f^{-1}_{\rm bias}$ is the inverse of the
monotonic function $f_{\rm bias}$.

\begin{figure}
\begin{center}
   \leavevmode \epsfxsize=16cm \epsfbox{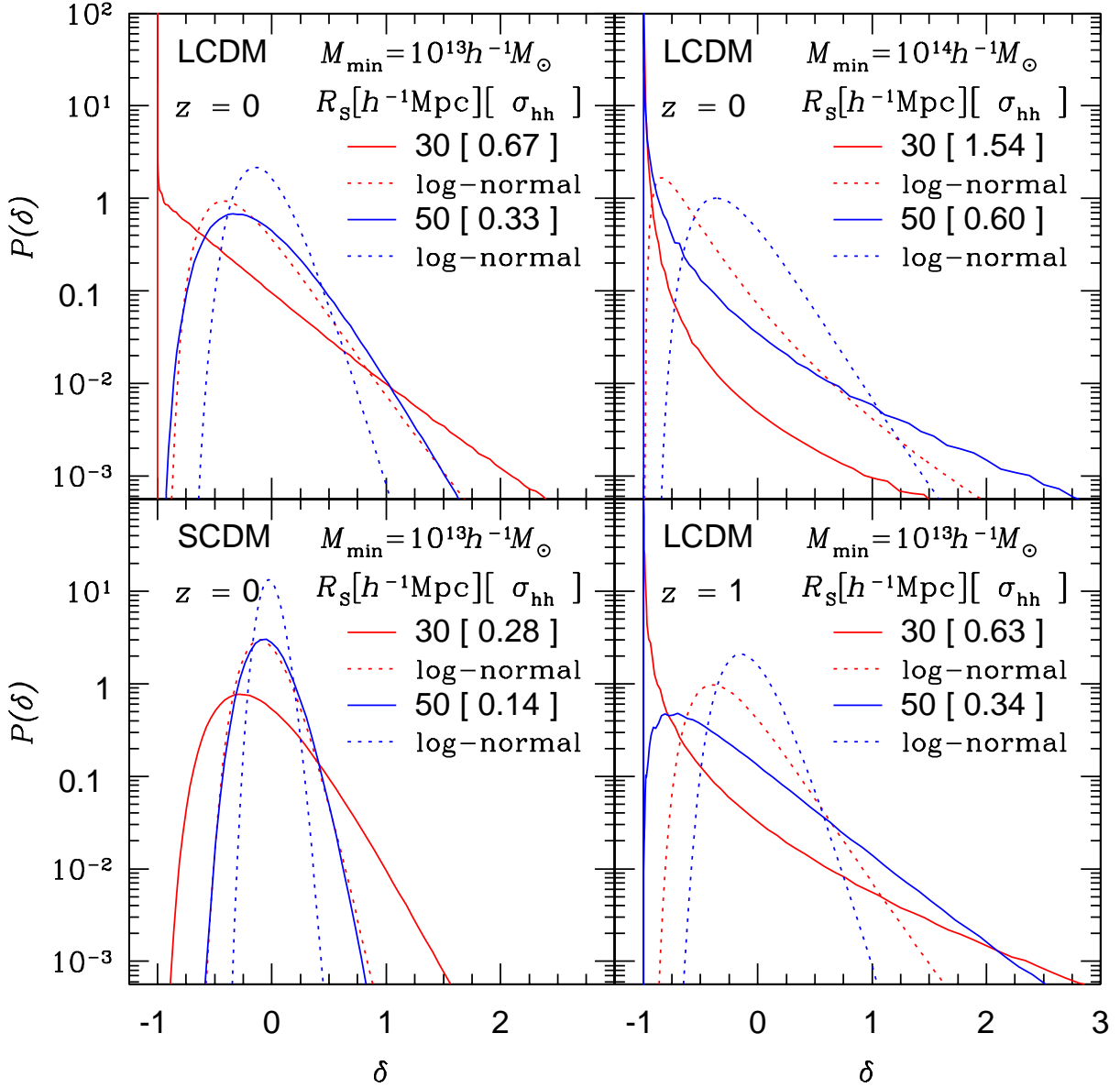}
\end{center}
\figcaption{ Probability distribution functions for dark halos with
$\Rs=30h^{-1}$Mpc and $50h^{-1}$Mpc.  For reference, dotted
lines indicate the log-normal distribution with the variance of the
corresponding $\sigmahh$.  {\it Upper-left:} LCDM model, $z=0$,
and $M_{\rm min} = 10^{13}h^{-1}M_{\odot}$ ; {\it Upper-right:} LCDM
model, $z=0$, and $M_{\rm min} = 10^{14}h^{-1}M_{\odot}$ ; {\it
Lower-left:} SCDM model, $z=0$, and $M_{\rm min} =
10^{13}h^{-1}M_{\odot}$ ; {\it Lower-right:} LCDM model, $z=1$, and
$M_{\rm min} = 10^{13}h^{-1}M_{\odot}$.  In all panels $M_{\rm max} =
10^{16}h^{-1}M_{\odot}$.
\label{fig:pdf_dh} }
\end{figure}

Figure \ref{fig:pdf_dh} plots the PDF for dark halos in solid lines,
respectively, employing the same sets of parameters as Figure
\ref{fig:biasing}. In order to separate the changes of the PDF due to
the nonlinear biasing and the nonlinear gravitational evolution, we
compare the dark halo PDF with the log-normal PDF
(eq:[\ref{eq:pdf_mass}]; dotted lines).  Also for that purpose the
variance $\sigmamm$ of the log-normal in equation (\ref{eq:pdf_mass}) is
adjusted to the corresponding $\sigmahh$ of the halos:
%%%%%%%%%%%%%%%%%%%%%%%%%%%%%%%%%%%%%%%%%%%%%%%%%%%%%%%%%%%%
\begin{equation}
\label{eq:sigmah}
\sigmahh^{2} \equiv \int\int
\deltah^{2}P(\deltam,\deltah)d\deltam d\deltah ,
\end{equation}
%%%%%%%%%%%%%%%%%%%%%%%%%%%%%%%%%%%%%%%%%%%%%%%%%%%%%%%%%%%
in plotting the log-normal PDF curves, Note that the quantity
$\sigma_{\rm hh}$ is somewhat different from variance of the averaged
biasing, $\langle[\overline\deltah(\deltam)]^2\rangle$ and contains the
nonlinearity and the stochasticity of biasing.  \footnote{ According to
Taruya \& Suto (2000), quantities $\sigma_{\rm hh}$ and
$\langle[\overline\deltah(\deltam)]^2\rangle$ are related in the
following way:
%%%%%%%%%%%%%%%%%%%%%%%%%%%%%%%%%%%%%%%%%%%%%%%%%%%%%%%%%%%%%%%
\begin{eqnarray}
\langle[\overline\deltah(\deltam)]^2\rangle = \sigmahh^2
	\,\,\,
\frac{1+\epsilon_{\rm nl}^2}{1+\epsilon_{\rm nl}^2+\epsilon_{\rm scatt}^2}, 
\nonumber
\end{eqnarray}
%%%%%%%%%%%%%%%%%%%%%%%%%%%%%%%%%%%%%%%%%%%%%%%%%%%%%%%%%%%%%%%
where $\epsilon_{\rm nl}$ and $\epsilon_{\rm scatt}$ respectively denote 
the degree of nonlinearity and stochasticity.
} 

Figure 2 shows the sensitivity of PDFs for the halo mass $M_{\rm min}$
and the cosmological parameters, especially for the fluctuation
amplitude $\sigma_8$.  The dark halo PDF significantly differs from the
log-normal PDF even using the same value of variance, which clearly
indicates the importance of the nonlinearity in the halo biasing.  In
some models, the PDF at $\deltah=-1$ diverges since $\deltah$ in our
model reaches $-1$ even though $\deltam$ is greater than $-1$.  Then
$f_{\rm bias}$ is not a one-to-one mapping between $\deltah$ and
$\deltam$ and the expression (\ref{eq:pdf_halo}) becomes singular there.
Of course, this divergence occurs only at $\deltah=-1$, and does not
affect our results for $\deltah>-1$.

\subsection{Genus Curve For Dark Halos}

As we describe in the previous subsection, our halo biasing model
results in the one-to-one mapping between $\deltaRG$ and
$\deltah=f_{\rm bias}(\deltam)=f_{\rm bias}(f_{\rm ln}(\deltaRG))$. 
Then using the prescription of Matsubara \& Yokoyama (1996),
we can derive the genus density explicitly.

For the nonlinear mass density $\deltam$ with the log-normal PDF
(\ref{eq:pdf_mass}), the genus density is computed as (Matsubara \&
Yokoyama 1996):
%%%%%%%%%%%%%%%%%%%%%%%%%%%%%%%%%%%%%%%%%%%%%%%%%%%%
\begin{eqnarray}
\label{eq:G_mass}
  g_{\rm mass}(\nu_{\rm m}) & =
  & \gmax ~ \exp\left[-\frac{\{\ln[(1+
  \nu_{\rm m}\sigmamm)\sqrt{1+\sigmamm^{2}}]\}^{2}}
  {2\ln(1+\sigmamm^{2})}\right] \nonumber \\
  & \times & \left[1-\frac{\{\ln[(1+\nu_{\rm m}\sigmamm)
  \sqrt{1+\sigmamm^{2}}]\}^{2}}{\ln(1+\sigmamm^{2})}\right] ,
\end{eqnarray}
%%%%%%%%%%%%%%%%%%%%%%%%%%%%%%%%%%%%%%%%%%%%%%%%%%%%%%%%%%%%%%%
where the level of isodensity contour $\nu_{\rm m}$ is given by 
$\delta_{\rm mass}/\sigma_{\rm mm}$. The quantity 
$\gmax$ is the maximum value of the genus:
%%%%%%%%%%%%%%%%%%%%%%%%%%%%%%%%%%%%%%%%%%%%%%%%%%%%%%%%%%%%%%
\begin{equation}
\label{eq:gmax}
  \gmax = g_{\rm mass}(\delta_{\rm\scriptscriptstyle MAX}/\sigmamm)
= \frac{1}{(2\pi)^{2}}\frac{\sigma^{3}_{\rm 1,m}}
  {[3(1+\sigmamm^{2})
  \ln(1+\sigmamm^{2})]^{3/2}} ,
\end{equation}
%%%%%%%%%%%%%%%%%%%%%%%%%%%%%%%%%%%%%%%%%%%%%%%%%%%%%%%%%%%%%%%
at $\delta_{\rm\scriptscriptstyle MAX} \equiv 1/\sqrt{1+\sigmamm^{2}} -1$ with
%%%%%%%%%%%%%%%%%%%%%%%%%%%%%%%%%%%%%%%%%%%%%%%%%%%%%%%%%
\begin{equation}
 \label{eq:sigmam1}
  \sigma_{\rm 1,m}^{2}(\Rs,z)=\int\frac{dk}{2\pi^{2}}
   k^{4}P_{\rm nl}(k,z)W^{2}(k\Rs)~ .
\end{equation}
%%%%%%%%%%%%%%%%%%%%%%%%%%%%%%%%%%%%%%%%%%%%%%%%%%%%%%%   

We transform this genus density using the mean biasing relation
$\deltah = f_{\rm bias}(\deltam)$.  Then, at a 
given threshold $\nu_{\rm h}\equiv\delta_{\rm halo}/\sigma_{\rm hh}$, 
our final expression for the genus density of dark halos becomes as follows:
%%%%%%%%%%%%%%%%%%%%%%%%%%%%%%%%%%%%%%%%%%%%%%%%%%%%
\begin{eqnarray}
\label{eq:G_halo}
  g_{\rm halo}(\nu_{\rm h}) & =
  & \gmax ~ \exp\left[-\frac{\{\ln[(1+f_{\rm bias}^{-1}
  (\nu_{\rm h}\sigmahh))\sqrt{1+\sigmamm^{2}}]\}^{2}}
  {2\ln(1+\sigmamm^{2})}\right] \nonumber \\
  & \times & \left[1-\frac{\{\ln[(1+f_{\rm bias}^{-1}
(\nu_{\rm h}\sigmahh))
  \sqrt{1+\sigmamm^{2}}]\}^{2}}{\ln(1+\sigmamm^{2})}\right] ,
\end{eqnarray}
%%%%%%%%%%%%%%%%%%%%%%%%%%%%%%%%%%%%%%%%%%%%%%%%%%%%%%%%%%%%%%%
with $\sigma_{\rm hh}$ defined by (15).  Here,
$\gmax=g_{\rm halo}[f_{\rm 
 bias}(\delta_{\rm\scriptscriptstyle MAX})/\sigmahh] 
 =g_{\rm mass}(\delta_{\rm\scriptscriptstyle MAX}/\sigmamm) $ is given also by equation (\ref{eq:gmax}) and is listed in
Table \ref{tab:genus_amplitude} for some models.

%%%%%%%%%%%%%%%%%%%%%%%%%%%%%%%%%%%%%%%%%%%%%%%%%%%%%%%%%%%%%%%
\section{PREDICTIONS OF GENUS FOR HALOS IN CDM MODELS \label{sec:Results}}

In this section we present several model predictions of the genus
density for dark matter halos with particular emphasis on its parameter
dependence. Then we examine to which extent our predictions are
reproduced with  a perturbation model on the nonlinear gravitational 
growth and the biasing.

\subsection{Parameter Dependence of the Halo Genus}

\begin{figure}
\begin{center}
   \leavevmode \epsfxsize=16cm \epsfbox{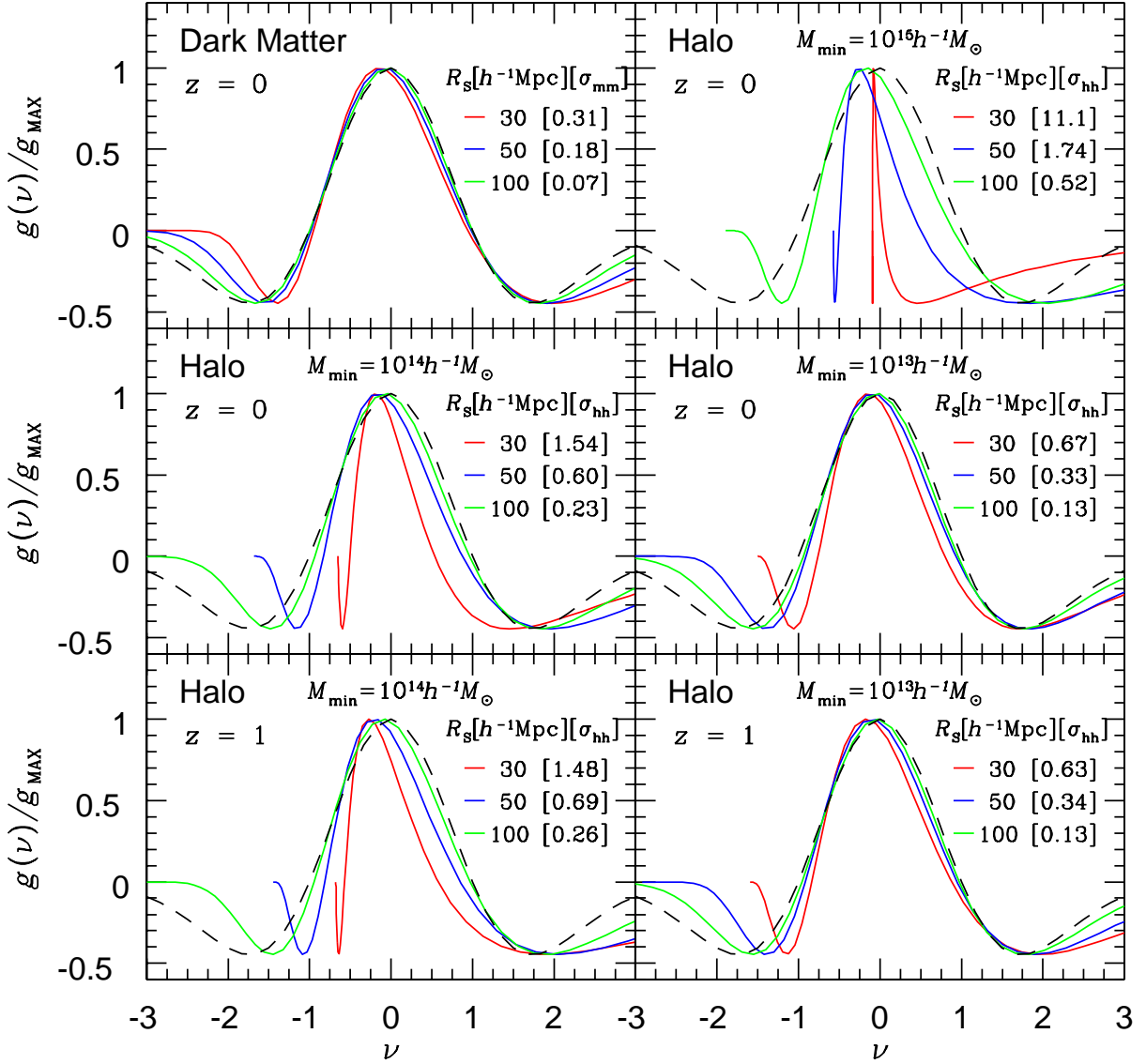}
\end{center}
\figcaption{ Normalized genus density $g(\nu)/\gmax$ 
for dark halos in LCDM model with $\Rs = 30h^{-1}$Mpc
(red), $50h^{-1}$Mpc (blue), and $100h^{-1}$Mpc (green).  For
comparison, {\it Upper-left} panel plots the normalized genus density
for dark matter.  The other panels assume ${M_{\rm
max}=10^{16}h^{-1}M_{\odot}}$.  {\it Upper-right:} $z=0$ and ${M_{\rm
min}=10^{15}h^{-1}M_{\odot}}$; {\it Middle-left:} $z=0$ and ${M_{\rm
min}=10^{14}h^{-1}M_{\odot}}$; {\it Middle-right:} $z=0$ and ${M_{\rm
min}=10^{13}h^{-1}M_{\odot}}$; {\it Lower-left:} $z=1$ and ${M_{\rm
min}=10^{14}h^{-1}M_{\odot}}$; {\it Lower-right:} $z=1$ and ${M_{\rm
min}=10^{13}h^{-1}M_{\odot}}$.
\label{fig:G_lcdm} }
\end{figure}

\begin{figure}
\begin{center}
   \leavevmode \epsfxsize=16cm \epsfbox{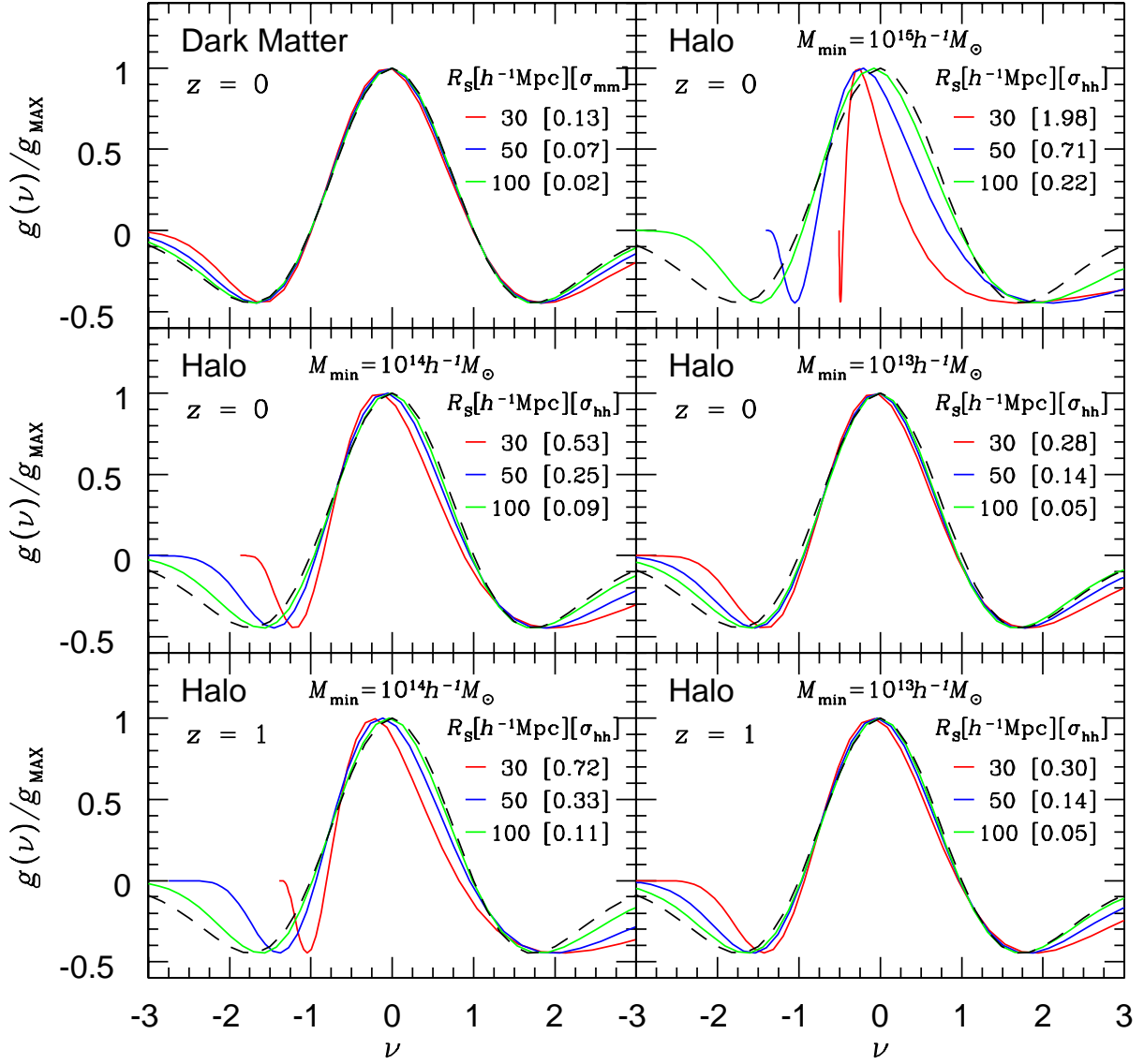}
\end{center}
\figcaption{ Same as Figure \ref{fig:G_lcdm} for SCDM model. 
\label{fig:G_scdm} }
\end{figure}

Consider the halo mass dependence first. As we argued in \S 2.1, the
maximum mass does not affect the results as long as $M_{\rm max}>
10^{16}M_\odot$. They are, however, very sensitive to the minimum mass
$M_{\rm min}$. This mass dependence is shown in Figures
\ref{fig:G_lcdm} and \ref{fig:G_scdm} in LCDM and SCDM models,
respectively, where we plot the {\it normalized} genus density (i.e.,
$g(\nu)$ in units of its maximum value $\gmax$). 
As $M_{\rm  min}$ increases, the nonlinearity of both the halo biasing and the
mass fluctuation amplitude becomes stronger (see Fig.\ref{fig:biasing}
and Table \ref{tab:biasing_fit}) which results in the significant
departure from the random-Gaussian prediction (dashed lines).

Also the genus is fairly dependent on the smoothing length $\Rs$
(Figs \ref{fig:G_lcdm} and \ref{fig:G_scdm}, and Table
\ref{tab:genus_amplitude}). While the genus with $\Rs \simgt
100\himpc$ approaches the random-Gaussian prediction, the deviation is
detectable with smaller $\Rs$. We discuss the detectability of
the signature in the last section.

In contrast, the redshift dependence is rather weak; neither the
amplitude and the shape of the genus density evolves much, at least
between $z=0$ and $1$ (which is the relevant range of the redshift for
galaxy clusters). This is interpreted as a result of the two competing
effects; the stronger clustering due to the halo biasing at higher $z$
and the weaker amplitude of 
the mass fluctuations(see also Taruya \& Yamamoto 2001).

The above features apply both to LCDM and SCDM.  The comparison of
Figures \ref{fig:G_lcdm} and \ref{fig:G_scdm} indicates that the
predictions in SCDM model is slightly closer to the random-Gaussian.
This is simply because these models are normalized according to the
cluster abundance and thus the mass fluctuation amplitude at $z=0$ is
smaller in SCDM ($\sigma_8=0.6$) than in LCDM ($\sigma_8=1.0$).

\subsection{Comparison with the Perturbation Analysis}

So far we have presented our predictions using the fully nonlinear
model of both the halo biasing and the mass fluctuations.  As long as
galaxy clusters are concerned, however, we are mainly interested in
fairly large $\Rs$. Therefore the perturbation analysis may be a
reasonable approximation. This line of studies is pioneered by
Matsubara (1994) in the weakly nonlinear evolution of mass density
field, and also by Matsubara (2000) including the effect of biasing.
We will perform the perturbation analysis of our full model
predictions. This attempt is instructive in understanding the origin
of the departure from the random-Gaussianity. Also the resulting
perturbation formula becomes useful since it is readily applicable to
a wide range of the biasing models, in addition to our halo biasing,
once their second-order biasing coefficients are given.

Matsubara (1994) derives the analytical formula for the genus of the
mass density field up to the lowest-order term
in the (linear) mass variance $\sigma$ :
%%%%%%%%%%%%%%%%%%%%%%%%%%%%%%%%%%%%%%%%%%%%%%%%%%%%%%%%%%%%%%
\begin{equation}
  \label{eq:second-order perturbation}
\hspace*{-0.5cm}
  g(\nu) \sim  -\frac{1}{(2\pi)^{2}}\left(\frac{\sigma_1}{\sqrt{3}
   \sigma}\right)^{3/2}
   \exp\left(-\frac{\nu^2}{2}\right)
   \left\{H_{2}(\nu)+\left[\frac{S^{(0)}}{6}H_5(\nu)
   +S^{(1)}H_3(\nu)+S^{(2)}H_1(\nu)\right]\sigma\right\},
\end{equation}
%%%%%%%%%%%%%%%%%%%%%%%%%%%%%%%%%%%%%%%%%%%%%%%%%%%%%%%%%%%%% 
where $S^{(a)}$ represents the skewness parameters defined by
%%%%%%%%%%%%%%%%%%%%%%%%%%%%%%%%%%%%%%%%%%%%%%%%%%%%%%%%%%%%
\begin{eqnarray}
   S^{(0)} & = & \frac{\langle\nu^{3}\rangle}{\sigma}, \\ 
   S^{(1)} & = & -\frac{3}{4}\frac{\langle\nu^{2}(\nabla^{2}\nu)\rangle
   \sigma}{\sigma_{1}^{2}}, \\
   S^{(2)} & = & -\frac{9}{4}\frac{\langle(\nabla\nu\cdot\nabla\nu)
   (\nabla^2\nu)\rangle\sigma^3}{\sigma_1^4}.
\end{eqnarray}
%%%%%%%%%%%%%%%%%%%%%%%%%%%%%%%%%%%%%%%%%%%%%%%%%%%%%%%%%%%%%

We expand equation (\ref{eq:G_mass}) with respect to $\sigmamm$
and find that it reduces to
%%%%%%%%%%%%%%%%%%%%%%%%%%%%%%%%%%%%%%%%%%%%%%%%%%%%%%%%%%%%%%%%%%%%
\begin{eqnarray}
  \label{eq:mass_pb}
   g_{\rm mass}(\nu_{\rm m}) = 
 g_{\rm\scriptscriptstyle RG}(\nu_{\rm m})\left\{1 +
  \frac{1}{2}H_3(\nu_{\rm m})\sigmamm 
   + {\cal O}(\sigmamm^2)\right\}, 
\end{eqnarray}
%%%%%%%%%%%%%%%%%%%%%%%%%%%%%%%%%%%%%%%%%%%%%%%%%%%%%%%%%%%%%%%%%%%%
where
%%%%%%%%%%%%%%%%%%%%%%%%%%%%%%%%%%%%%%%%%%%%%%%%%%%%%%%%%%%%%%%%%%%%%%%%
\begin{eqnarray}
   g_{\rm\scriptscriptstyle RG}(\nu_{\rm m}) =  \frac{1}{(2\pi)^{2}} 
   \left(\frac{\sigma^{2}_{\rm 1,m}}{3\sigma^{2}_{\rm m}}\right)^{3/2}
   \exp\left(-\frac{\nu_{\rm m}^2}{2}\right)(1-\nu_{\rm m}^{2}).
\end{eqnarray}
%%%%%%%%%%%%%%%%%%%%%%%%%%%%%%%%%%%%%%%%%%%%%%%%%%%%%%%%%%%%%%%%%%%%%%%%
This implies that all the above skewness parameters are equal to 3 in
the log-normal PDF.

If we further include the halo biasing effect using the second-order
fit $\deltah=f_{\rm bias}^{\rm fit(2)}(\deltam)$ given by equation
(\ref{eq:biasing_fit}), we obtain the perturbation formula of our halo
genus up to $\sigmahh$ as
%%%%%%%%%%%%%%%%%%%%%%%%%%%%%%%%%%%%%%%%%%%%%%%%%%%%%%%%%%%%
\begin{eqnarray}
\label{eq:genus_pb}
   g_{\rm halo}(\nu_{\rm h}) =  g_{\rm halo,0}(\nu_{\rm h})\left\{1+
   \frac{1}{2}\frac{b_1+b_2}{b_1 b_{\rm var}}H_3\left(\frac{b_{\rm var}
   \nu_{\rm h}}{b_1}\right)\sigmahh + 
   {\cal O}(\sigmahh^2)\right\} , 
\end{eqnarray}
%%%%%%%%%%%%%%%%%%%%%%%%%%%%%%%%%%%%%%%%%%%%%%%%%%%%%%%%%%%%
where
%%%%%%%%%%%%%%%%%%%%%%%%%%%%%%%%%%%%%%%%%%%%%%%%%%%%%%%%%%%%
\begin{eqnarray}
   g_{\rm halo,0}(\nu_{\rm h})  =   \frac{1}{(2\pi)^{2}} 
   \left(\frac{\sigma^{2}_{\rm 1,m}}{3\sigma^{2}_{\rm m}}\right)^{3/2}
   \exp\left\{-\frac{1}{2}\left(\frac{b_{\rm 
    var}\nu_{\rm h}}{b_1}\right)^2\right\}
   \left\{1-\left(\frac{b_{\rm var}\nu_{\rm h}}{b_1}\right)^2\right\} ,
\end{eqnarray}
%%%%%%%%%%%%%%%%%%%%%%%%%%%%%%%%%%%%%%%%%%%%%%%%%%%%%%%%%
with $b_{\rm var} = \sigmahh/\sigmamm$ (Table
\ref{tab:biasing_fit}).

If the biasing is linear and deterministic, $b_{\rm var} = b_1$ and the
skewness parameters reduce to
%%%%%%%%%%%%%%%%%%%%%%%%%%%%%%%%%%%%%%%%%%%%%%%%%%%%%%%
\begin{equation}
   S_{\rm h}^{(0)} = S_{\rm h}^{(1)} = S_{\rm h}^{(2)} =
  \frac{3}{b_1}\left(1+\frac{b_2}{b_1}\right) ,
\end{equation}
%%%%%%%%%%%%%%%%%%%%%%%%%%%%%%%%%%%%%%%%%%%%%%%%%%%%%%%%%
which agrees with equation [4.82] of Matsubara (2000).

\begin{figure}
\begin{center}
   \leavevmode \epsfxsize=16cm \epsfbox{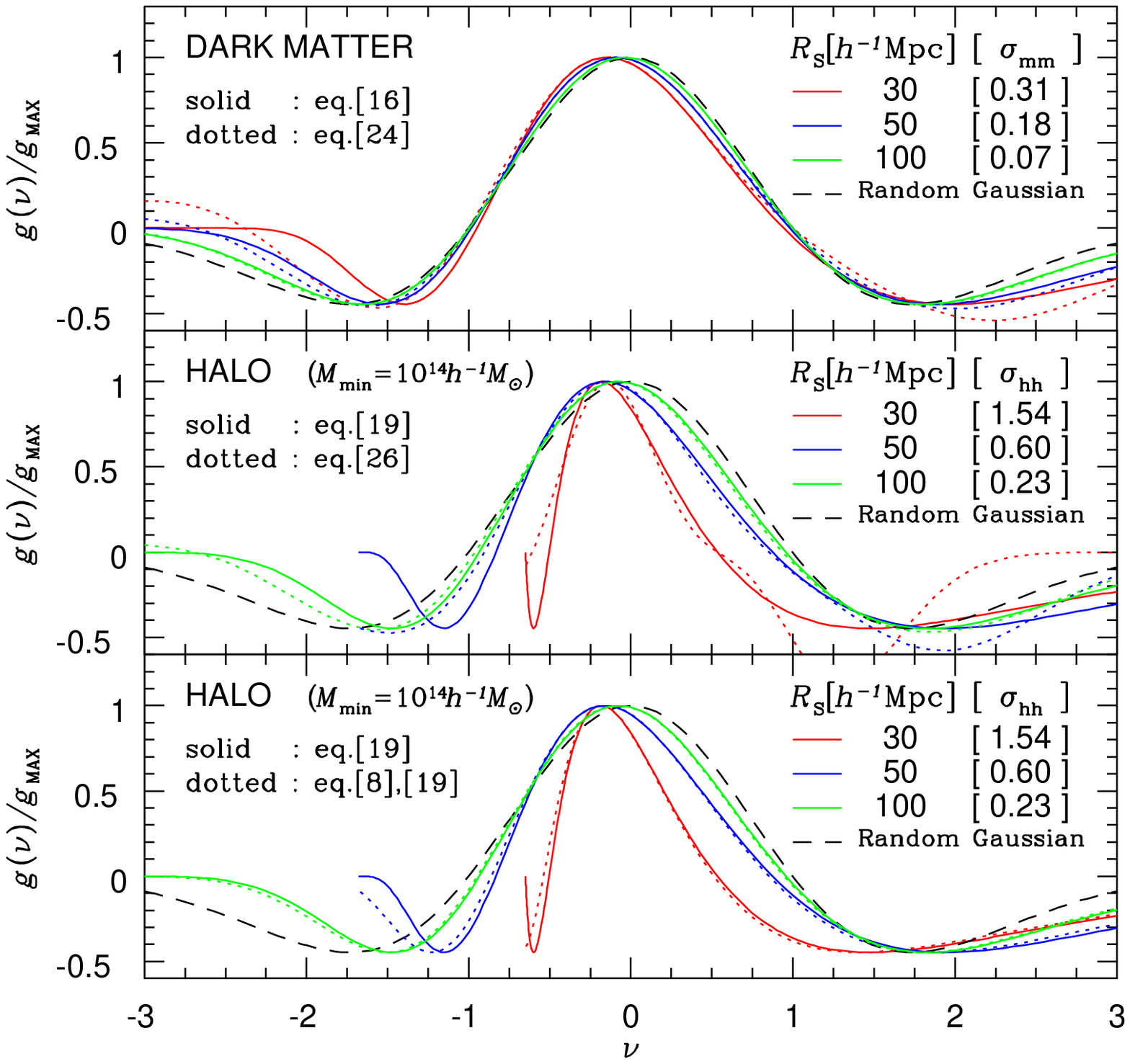}
\end{center}
\figcaption{Comparison of $g(\nu)/\gmax$
between the full and perturbation models. The predictions assume $z =
0$ in LCDM model, $M_{\rm min}=10^{14}h^{-1}M_{\odot}$, $M_{\rm
max}=10^{16}h^{-1}M_{\odot}$, and three different smoothing scales;
$\Rs=30h^{-1}$Mpc (red), $50h^{-1}$Mpc (blue), and
$100h^{-1}$Mpc (green).  {\it Top:} mass in the log-normal model
(solid; eq.[\ref{eq:G_mass}]) and in the perturbation model (dotted;
eq.[\ref{eq:mass_pb}]).  {\it Middle:} dark halos (solid;
eq.[\ref{eq:G_halo}]) and the perturbation formula of our halo genus
with the second-order biasing model (dotted; eq.[\ref{eq:genus_pb}]).
{\it Bottom:} dark halos (solid; eq.[\ref{eq:G_halo}]) and the results
combining the log-normal mass PDF and the second-order biasing
model(dotted; eqs.[\ref{eq:biasing_fit}], [\ref{eq:G_halo}]).
\label{fig:perturb_genus}}
\end{figure}

Figure {\ref{fig:perturb_genus}} compares those perturbation results
with our full model predictions; {\it Top} panel shows the normalized
genus in the log-normal model (solid lines; eq.[\ref{eq:G_mass}]) and
in the perturbation model (dotted lines; eq.[\ref{eq:mass_pb}]).
Dotted lines in {\it Middle} panel refer to the results combining 
the weakly nonlinear evolution model (eq.[\ref{eq:mass_pb}])
and the second-order biasing model (eq.[\ref{eq:biasing_fit}]), 
while in {\it Bottom} panel they refer to the results combining 
the log-normal mass PDF (eq.[\ref{eq:G_mass}]) and the second-order biasing
model (eq.[\ref{eq:biasing_fit}]). Those are to be compared with our
full predictions (solid lines). 
In the perturbation results (eqs.[\ref{eq:mass_pb}], [\ref{eq:genus_pb}])
we define $\gmax$ respectively
as the maximum value in the set of the plotted genus respectively.
This comparison indicates that the
second-order biasing model provides a very good approximation for the
parameters of interest, also that the lowest-order correction to the
nonlinear mass evolution dominates for large smoothing lengths 
$\Rs \simgt 50\himpc$. Therefore we conclude that our perturbation
formula (\ref{eq:genus_pb}) is in practice a useful and reliable
approximation to the genus of galaxy halos, which is applicable even
beyond our particular biasing model discussed here.

%%%%%%%%%%%%%%%%%%%%%%%%%%%%%%%%%%%%%%%%%%%%%%%%%%%%%%%%%%%%%%%
\section{SUMMARY AND DISCUSSION}

In the present paper, we have presented an analytic model of genus for
dark matter halos adopting the stochastic nonlinear halo biasing model
by Taruya \& Suto (2000), assuming that the primordial mass density
field obeys the random-Gaussian statistics.  This is the first attempt
to predict the genus statistics simultaneously taking account of the
nonlinear nature in the biasing and of the nonlinear
gravitational evolution of the underlying mass distribution.  In the
remainder of this section, we discuss several implications of our model
predictions.

First of all, we note that our model applies only for dark matter
halos identified according to the spherical collapse model strictly
speaking. Nevertheless this picture is now widely accepted as an
empirical model for galaxy clusters in the universe. In fact, the
excellent agreement between the predicted halo abundance and the
observed cluster abundance (e.g., Kitayama \& Suto 1997; Suto 2000 and
references therein) justifies the empirical one-to-one correspondence
between the theoretical halo and the observed cluster in a statistical
sense. Adopting this conventional view, our predictions for dark
matter halos can be readily translated into those for clusters on the
basis of the mass--temperature and mass-luminosity relations at the
present epoch, for instance, as
%%%%%%%%%%%%%%%%%%%%%%%%%%%%%%%%%%%%%%%%%%%%%%%%%%%%%%%%%%%
\begin{equation}
\label{eq:TM}
k_{\rm\scriptscriptstyle B}T 
= 1.3\left(\frac{\Delta_{\rm c}}{18\pi^2}\right)^{1/3}
\left(\frac{M}{10^{14}h^{-1}M_{\odot}}\right)^{2/3}
\Omega_0^{1/3}{\rm keV} ,
\end{equation}
%%%%%%%%%%%%%%%%%%%%%%%%%%%%%%%%%%%%%%%%%%%%%%%%%%%%%%%%%%%
and
%%%%%%%%%%%%%%%%%%%%%%%%%%%%%%%%%%%%%%%%%%%%%%%%%%%%%%%%%
\begin{equation}
\label{eq:LM}
L_{\rm bol} = 1.7\left(\frac{\Delta_{\rm c}}{18\pi^2}\right)^{1.1}
\left(\frac{M}{10^{14}h^{-1}M_{\odot}}\right)^{2.3}
\Omega_0^{1.1}10^{42}h^{-2}{\rm erg~sec^{-1}},
\end{equation}
%%%%%%%%%%%%%%%%%%%%%%%%%%%%%%%%%%%%%%%%%%%%%%%%%%%%%%%%%
where $k_{\rm\scriptscriptstyle B}$ 
is Boltzmann constant and $\Delta_{\rm c}$ is the
mean density of a virialized cluster (Kitayama \& Suto 1996; Suto et
al. 2000). Therefore our model predictions can be tested against
future cluster observations in various wavebands.

Next it should be stressed that our current model is based on the
one-to-one mapping of the primordial density field that obeys the
random-Gaussian statistics. In this approximation, if we express our
genus predictions as a function of $\overline{\nu}$ defined through
the volume fraction $f$:
%%%%%%%%%%%%%%%%%%%%%%%%%%%%%%%%%%%%%%%%%%%%%%%%%%%%%%%%%%%%%%%
\begin{equation}
\label{eq:volfra}
f=\frac{1}{\sqrt{2\pi}}\int^{\infty}_{\overline{\nu}}
\exp\left(-\frac{t^2}{2}\right)dt ,
\end{equation}
%%%%%%%%%%%%%%%%%%%%%%%%%%%%%%%%%%%%%%%%%%%%%%%%%%%%%%%%%%%%%%%%
they simply reduce to the random-Gaussian prediction
(eq.[\ref{eq:genus_rd}]).  Most previous studies used this
$\overline{\nu}$ in order to remove the (unknown) distortion of the PDF
due to the gravitational nonlinear evolution.  Nevertheless we prefer to
adopt the {\it correct} definition of $\nu =\delta/\sigma$ because the
departure from the Gaussian prediction (eq.[\ref{eq:genus_rd}]) provides
also an important measure of the gravitational nonlinear evolution and
the nonlinear biasing in more advanced approach than what we took here.
For instance, the perturbative results in weakly nonlinear regime
(Matsubara 1994) do not reduce to equation (\ref{eq:genus_rd}) even in
terms of $\overline{\nu}$, and also the effect of stochasticity in
biasing, which is ignored in the present analysis, would add another
feature which violates the simple scaling idea behind the use of
$\overline{\nu}$.  As we have shown in the present paper, the genus as a
function $\nu$ can be a useful probe of the nonlinear gravitational
evolution and the biasing of objects and mass, as long as the
random-Gaussianity of the primordial density field is correct.  The
comparison between $g(\overline{\nu})$ and $g(\nu)$ is discussed in
detail by Colley et al. (2000).

In our genus model, the overall amplitude of genus for dark halos
$\gmax$ depends only on mass density spectrum.
Therefore the observed amplitude of the cluster genus may provide direct
information on the mass fluctuation in principle.  However this fact is
heavily dependent on our assumption of one-to-one mapping of density
fluctuation in equation (\ref{eq:mean_biasing}), and needs to be checked
with future numerical simulations.

\begin{figure}
\begin{center}
   \leavevmode \epsfxsize=16cm \epsfbox{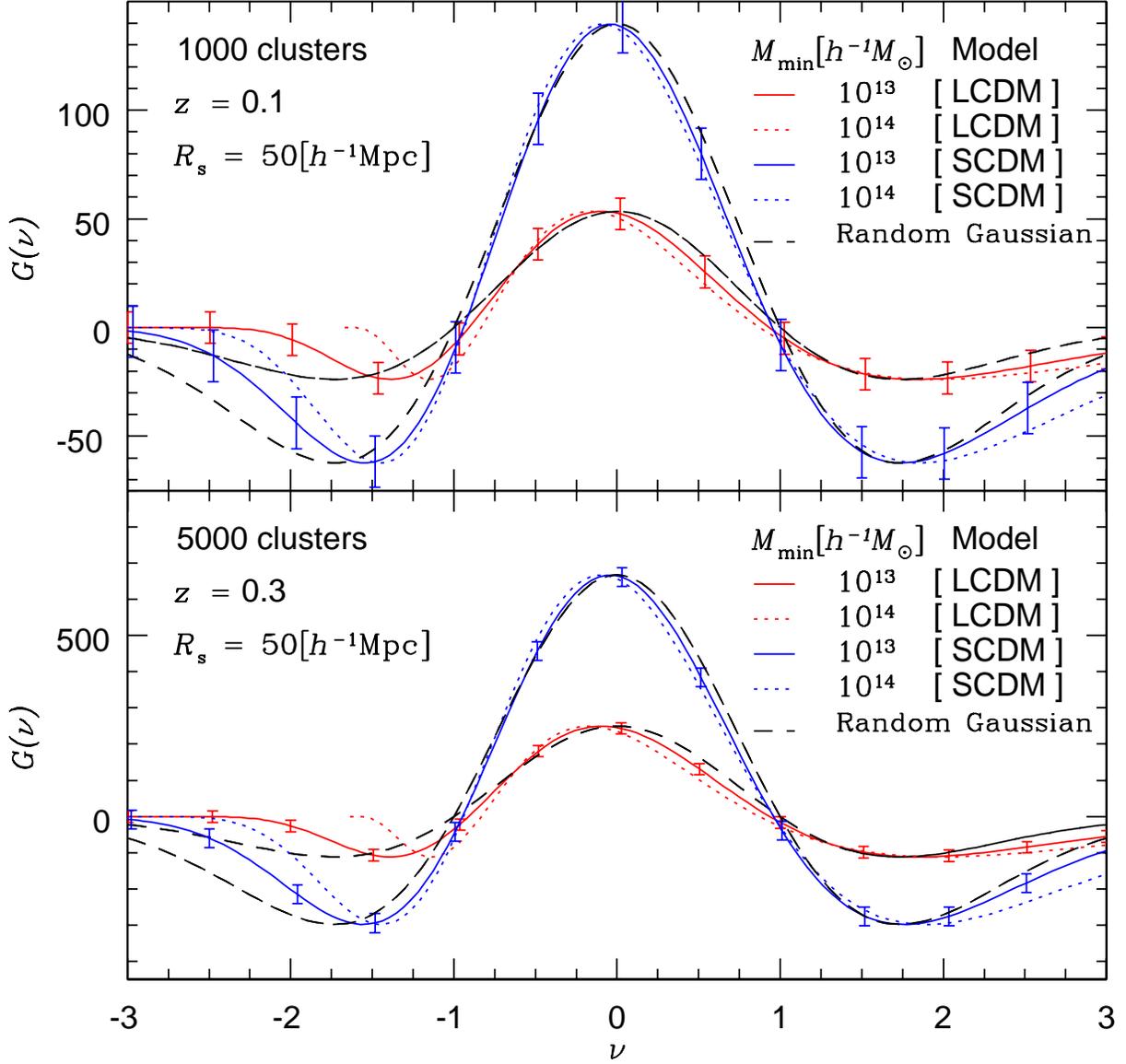}
\end{center}
\figcaption{ Expected total number of genus for surveys corresponding to
 $N \simeq 1000$ ({\it Upper} panel) and $N \simeq 5000$ ({\it Lower}
 panel); $M_{\rm min}=10^{13}h^{-1}M_{\odot}$ (solid) and $M_{\rm
 min}=10^{14}h^{-1}M_{\odot}$ (dotted) in LCDM (red) and SCDM (blue)
 models. All the curves adopt top-hat smoothing of $\Rs=50\himpc$.
 \label{fig:error_genus} }
\end{figure}

Finally we briefly discuss the future detectability of the departure of
the halo genus from the random-Gaussian prediction.  We use the top-hat
smoothing $\Rs = 50h^{-1}$ Mpc and compute our model predictions for
$M_{\rm min}=10^{13}h^{-1}M_\odot$ and $10^{14}h^{-1}M_\odot$ in SCDM
and LCDM models.  Figure \ref{fig:error_genus} plots the total number of
genus $G(\nu)$; {\it Upper} and {\it Lower} panels are normalized so as
to roughly correspond to the Sloan Digital Sky Survey (SDSS)
spectroscopic and photometric cluster samples, respectively.  We do not
take into account the light cone effect (Yamamoto \& Suto 1999; Suto et
al. 1999) but rather show the results at the mean redshift $z=0.1$ and
$z=0.3$.  Taruya \& Yamamoto (2001) find that the light-cone effect is
not so significant for genus statistics up to $z \sim 1$.  The SDSS is
expected to detect $N_{\rm spect} \sim 1000$ clusters from spectroscopic
data up to $z \sim 0.2$ and $N_{\rm photo} \sim 5000$ from photometric
data up to $z \sim 1$.  Assuming that errors of genus are dominated by
the sampling noise, we assign the error independently of $\nu$ as
$\delta G = \sqrt{\gmax V_{\rm eff}}$, where
$V_{\rm eff}$ is the effective volume of the sample.  We further assume
that $V_{\rm eff}$ is proportional to the number of clusters.  If we
adopt the value for the Abell cluster sample of Rhoads et al.(1994),
$\sim 250$ Abell clusters in $V_{\rm eff} =9.5 \times 10^{6}
\,\,(\himpc)^{3}$, $V_{\rm eff}$ for SDSS spectroscopic and photometric
samples is estimated to be 4 and 20 times larger than that of Rhoads et
al.(1994), respectively.  The amplitude of the error bars estimated in
this way indicates that the departure from the random-Gaussianity should
be indeed detectable especially in photometric redshift data and that
one can even discriminate among some model predictions with the upcoming
data of galaxy clusters.

We plan to improve our model predictions by including the stochasticity
in the biasing model, the light-cone effect, and redshift-space
distortion. Also we are currently working on evaluating the genus of
halos identified from the cosmological N-body simulations. These results
and comparison will be reported elsewhere.

\acknowledgments

We thank the referee, Takahiko Matsubara, for a critical reading and
useful comments, and Wataru Kawasaki for discussion on the upcoming
cluster surveys.  A.T. gratefully acknowledges support from a JSPS
(Japan Society for the Promotion of Science) fellowship.  Numerical
computations were carried out at ADAC (the Astronomical Data Analysis
Center) of the National Astronomical Observatory, Japan and at KEK
(High Energy Accelerator Research Organization, Japan). This research
was supported in part by the Grant-in-Aid by the Ministry of
Education, Science, Sports and Culture of Japan (07CE2002, 12640231)
to RESCEU, and by the Supercomputer Project (No.99-52, N0.00-63) of
KEK.

\bigskip
%%%%%%%%%%%%%%%%%%%%%%%%%%%%%%%%%%%%%%%%%%%%%%%%%%%%%%%%%%%%%%%%%%%%%%%

%%%%%%%%%%%%%%%%%%%%%%%%%%%%%%%%%%%%%%%%%%%%%%%%%%%%%%%%%%%%%%%%%%%%%%%

\clearpage

%%%%%%%%%%%%%%%%%%%%%%%%%%%%%%%%%%%%%%%%%%%%%%%%%%%%%%%%%%%%%%%%%%%
\begin{deluxetable}{ccccccc}
\footnotesize
\tablecaption{The linear and second-order biasing coefficients,
$b_1$ and $b_2$(eq.[\ref{eq:biasing_fit}]), respectively, 
and the ratio of the rms fluctuations 
between dark halos and mass $b_{\rm var}$
($=\sigmahh/\sigmamm$)
for different cosmological models, $z$, $M_{\rm min}$, and $\Rs$.}
\tablewidth{0pt}
\tablehead{
  \colhead{Model} & \colhead{$z$}
  & \colhead{$M_{\rm min}[h^{-1}M_{\rm \odot}]$} 
  & \colhead{$\Rs[h^{-1}{\rm Mpc}]$} & \colhead{$b_1$} 
  & \colhead{$b_2$} & \colhead{$b_{\rm var}$}
}
\startdata
  LCDM & 0 & $10^{13}$ & 30  & 1.77 & 0.23 & 2.12 \\ 
  LCDM & 0 & $10^{13}$ & 50  & 1.76 & 0.20 & 1.87 \\ 
  LCDM & 0 & $10^{13}$ & 100 & 1.76 & 0.14 & 1.84 \\ 
  LCDM & 0 & $10^{14}$ & 30  & 2.72 & 3.26 & 4.90 \\ 
  LCDM & 0 & $10^{14}$ & 50  & 2.92 & 3.20 & 3.34 \\ 
  LCDM & 0 & $10^{14}$ & 100  & 3.00 & 3.70 & 3.29 \\ 
  SCDM & 0 & $10^{13}$ & 30  & 1.94 & 0.51 & 2.07 \\ 
  SCDM & 0 & $10^{13}$ & 50  & 1.94 & 0.77 & 2.04 \\ 
  LCDM & 1 & $10^{13}$ & 30  & 2.85 & 2.78 & 3.22 \\ 
  LCDM & 1 & $10^{13}$ & 50  & 2.90 & 2.60 & 3.08 \\ 
\enddata
\label{tab:biasing_fit}
\end{deluxetable}
%%%%%%%%%%%%%%%%%%%%%%%%%%%%%%%%%%%%%%%%%%%%%%%%%%%%%%%%%%%%%%%%%%%%%

%%%%%%%%%%%%%%%%%%%%%%%%%%%%%%%%%%%%%%%%%%%%%%%%%%%%%%%%%%%%%%%%%%%%%
\begin{deluxetable}{cccc}
\footnotesize
\tablecaption{The maximum values of the genus $\gmax$ 
for different cosmological models, $z$, and $\Rs$.
\label{tab:genus_amplitude}
}
\tablewidth{0pt}
\tablehead{
  \colhead{Model} & \colhead{$z$} 
  & \colhead{$\Rs[h^{-1}{\rm Mpc}]$} & 
\colhead{$\gmax$[$(100h^{-1}{\rm Mpc})^{-3}$]}
}
\startdata
  LCDM & 0 &  30 & 4.75  \\ 
  LCDM & 0 &  50 & 1.49  \\ 
  LCDM & 0 &  100 & 0.38  \\ 
  LCDM & 1 &  30 & 3.56  \\ 
  LCDM & 1 &  50 & 1.13  \\ 
  LCDM & 1 &  100 & 0.29  \\ 
  SCDM & 0 &  30 & 9.47  \\ 
  SCDM & 0 &  50 & 3.83  \\ 
  SCDM & 0 &  100 & 1.48  \\ 
  SCDM & 1 &  30 & 7.66  \\ 
  SCDM & 1 &  50 &  3.12 \\ 
  SCDM & 1 &  100 &  1.22 \\ 
\enddata
\end{deluxetable}
%%%%%%%%%%%%%%%%%%%%%%%%%%%%%%%%%%%%%%%%%%%%%%%%%%%%%%%%%%%%%%%%%%%%%

%%%%%%%%%%%%%%%%%%%%%%%%%%%%%%%%%%%%%%%%%%%%%%%%%%%%%%%%%%%%%%%%%%%%%
%%%%%%%%%%%%%%%%%%%%%%%%%%%%%%%%%%%%%%%%%%%%%%%%%%%%%%%%%%%%%%%%%%%%%

%%%%%%%%%%%%%%%%%%%%%%%%%%%%%%%%%%%%%%%%%%%%%%%%%%%%%%%%%%%%%%%%%%%%%
%%%%%%%%%%%%%%%%%%%%%%%%%%%%%%%%%%%%%%%%%%%%%%%%%%%%%%%%%%%%%%%%%%%

%%%%%%%%%%%%%%%%%%%%%%%%%%%%%%%%%%%%%%%%%%%%%%%%%%%%%%%%%%%%%%%%%%%%%
%%%%%%%%%%%%%%%%%%%%%%%%%%%%%%%%%%%%%%%%%%%%%%%%%%%%%%%%%%%%%%%%%%%%%

%%%%%%%%%%%%%%%%%%%%%%%%%%%%%%%%%%%%%%%%%%%%%%%%%%%%%%%%%%%%%%%%%%%%%
%%%%%%%%%%%%%%%%%%%%%%%%%%%%%%%%%%%%%%%%%%%%%%%%%%%%%%%%%%%%%%%%%%%%%

%%%%%%%%%%%%%%%%%%%%%%%%%%%%%%%%%%%%%%%%%%%%%%%%%%%%%%%%%%%%%%%%%%%%%
%%%%%%%%%%%%%%%%%%%%%%%%%%%%%%%%%%%%%%%%%%%%%%%%%%%%%%%%%%%%%%%%%%%%%

%%%%%%%%%%%%%%%%%%%%%%%%%%%%%%%%%%%%%%%%%%%%%%%%%%%%%%%%%%%%%%%%%%%%%
%%%%%%%%%%%%%%%%%%%%%%%%%%%%%%%%%%%%%%%%%%%%%%%%%%%%%%%%%%%%%%%%%%%%%

\end{document}